\documentclass [12pt]{article}
\usepackage[english]{babel}
\if@twoside\oddsidemargin25mm
\else\oddsidemargin25mm\evensidemargin25mm\marginparwidth25mm\fi%
\begin{document}
\begin{titlepage}
\begin{flushright}
KUL-TF-99/44\\
hep-th/9912178
\end{flushright}

\vfill
\begin{center}
{\large\bf Point Splitting and \(U(1)\) Gauge Invariance}
\\ \vskip 27.mm
{\bf D.~Olivi\'e }\\ \vskip .5cm
Institute for Theoretical Physics, University of Leuven, B-3001 Leuven,
Belgium
\end{center}

\vskip 10.mm
\vfill
\begin{center}
{\bf Abstract}
\end{center}
\begin{quote}
\small
A gauge transformation in quantum electrodynamics involves the product of
field operators at the same space-time point and hence does not have a
well-defined meaning.  One way to avoid this difficulty is to generalize the
gauge transformation by using different space-time points in the spirit of
Dirac's point splitting.  Such a generalization indeed exists and the resulting
generalized infinitesimal gauge transformation takes the form of an infinite
series in the coupling constant.  In this text I will present two examples of
generalized gauge transformations.
\end{quote}
\vspace{2mm} \vfill
\normalsize
\end{titlepage}

\section{Introduction}
It is well known that, in local quantum field theory, one encounters
divergences which arise from taking products of field operators at the same
space-time point.  As a result, these products do not have a well-defined
meaning.  Quite some time ago, Dirac~\cite{Dirac34} 
suggested point splitting as a remedy
for this difficulty: instead of taking all the field operators at the
space-time point $x$, a fixed four-vector $\epsilon$ is introduced so that only
field operators with different arguments $(x,x \pm \epsilon, \ldots)$ appear in
their products.  As long as $\epsilon$ is taken to be different from 0, the
products of field operators are well defined and the theory can be expected to
be free of divergences, \mbox{i.e.}, to be regularized.  At the end of the
calculation, the limit $\epsilon \rightarrow 0$ is taken, in order to recover
the original theory.

{\it A prioiri}, there are many ways in which such a point splitting procedure
can be implemented.  For gauge theories, however, one must ensure that the
introduction
of this new parameter $\epsilon$ preserves the invariance under gauge
transformations.  It thus appears reasonable that one first attempts to
construct gauge transformations involving products of field operators taken at
different space-time points, which we shall call generalized gauge
transformations.  Once such generalized gauge transformations are found, one can
then attempt to construct an action, which is invariant under these generalized 
gauge
transformations.

In this paper, I shall show how these ideas can be put to work for the \(U(1)\)
gauge symmetry, although several aspects can be carried over to the general
Yang-Mills case~\cite{Gastmans96,Gastmans98}.

\section{Framework}

The standard infinitesimal gauge transformations, $\delta_\Lambda$, for the
photon field \(A_\mu(x)\) and the Dirac field $\psi(x)$ in the Abelian
\(U(1)\) case take the form
\begin{eqnarray} 
\delta_\Lambda \, A_\mu(x)& = &- \partial_\mu \, \Lambda(x) \, ,
\nonumber\\
\delta_\Lambda \, \psi(x) &= &- \imath \, e \, \Lambda(x) \, \psi(x) 
\, , \label{eq2}\\
\delta_\Lambda \, \bar\psi(x) &=& \imath \, e \, \Lambda(x) \, \bar\psi(x) 
\, ,
\nonumber \end{eqnarray}
where $\Lambda(x)$ is the gauge parameter.
The gauge transformations of the fields satisfy the \(U(1)\) group
property: the commutator of two subsequent gauge transformations vanishes. 

In what follows, I shall require that the generalized gauge transformations
preserve this Abelian character, \mbox{i.e.}, two such generalized transformations
should commute.  This requirement imposes strong restrictions on the form such 
transformations
can take. Nevertheless, it was found that such infinitesimal transformations
can be constructed and that, for gauge transformations on the fermion fields,
they take the form of an infinite series in powers of the coupling constant
$e$:
\begin{equation}
\delta_\Lambda \, \psi(x) = \sum\limits_{n = 1}^{\infty}
\, e^n \, \delta_\Lambda^{(n)} \, \psi(x) \, . 
\end{equation}
This is to be contrasted  with the standard \(U(1)\) case
without point splitting, where the {\it finite} gauge transformations on the
fermion field are of infinite order in the coupling constant, the {\it
infinitesimal} ones in Eqs.\,(\ref{eq2}) being only of first order in $e$.

The action $A$, invariant under these infinitesimal generalized gauge
transformations,
is also an infinite series in the coupling constant $e$:
\begin{equation}
 A = \int d^4x \, {\cal L}(x) = \sum_{n = 0}^{\infty}\, e^n \, \int d^4x \,
 {\cal L}^{(n)}(x)
 \,,
\end{equation}
where ${\cal L}(x)$ is the Lagrangian density and ${\cal L}^{(n)}(x)$ its expansion in powers
of $e$.  The action being an infinite series in $e$
leads to new peculiar Feynman rules.  Besides the one-photon vertex, there 
are in this
generalized theory also two-, three-, four-, $\ldots$ photon vertices.

In Sections~3 and 4, I shall present two explicit examples of generalized gauge
transformations.

\section{First example}

Perhaps the simplest Ansatz one can imagine for generalized gauge
transformations is
\begin{eqnarray} 
\delta_\Lambda \, A_\mu(x) &=& - \partial_\mu \, \Lambda(x) \, ,
\nonumber \\
 \delta_\Lambda^{(1)} \, \psi(x) &=&  - \imath  
\, \Lambda(x + \epsilon) \, \psi(x + 2  \epsilon) \,,
\end{eqnarray}
where $\epsilon$ is the point splitting four-vector as discussed in the
introduction.
To satisfy the Abelian group property, one can take, \mbox{e.g.},
\begin{equation}
\delta_\Lambda^{(2)} \, \psi(x) = - \frac{1}{2} \, 
[\Lambda(x + \epsilon) + \Lambda(x + 3  \epsilon)]
 \, \psi(x +4\epsilon) \int_{x + \epsilon}^{x + 3 \epsilon} 
dy^{\alpha} \, A_\alpha(y),
\end{equation}
and so on for the higher order terms.
  
When $\epsilon \rightarrow 0$, then $\delta_\Lambda^{(1)} \, \psi(x)$
reduces to the standard gauge transformation~$[$Eqs.\,(\ref{eq2})$]$, and
$\delta_\Lambda^{(2)}\, \psi(x) \rightarrow 0$.
The general proof that the higher order terms 
\(\delta_\Lambda^{(n)} \, \psi(x)\) indeed exist, can be found in
Ref.\,\cite{Gastmans94}.

The expansion of the corresponding invariant action then yields the following
lowest order results for the Lagrangian density:
\begin{eqnarray}
{\cal L}^{(0)} &=& \bar\psi(x) \, (\imath \, 
\gamma^\mu \, 
\partial_\mu - m) \, \psi(x) - \frac{1}{4} \, F_{\mu \nu}(x) \, 
F^{\mu \nu}(x) \,, 
\\
{\cal L}^{(1)} &=& \bar\psi(x - \epsilon) \, \gamma^\mu  \,
\psi(x + \epsilon) \,  A_\mu(x)\,,
\\
{\cal L}^{(2)} &=& \frac{1}{2} \,  
\bar\psi(x - 2 \epsilon) 
\, \gamma^\mu \, \psi(x + 2 \epsilon) 
\nonumber \\
&& \times \bigg[[ A_\mu(x - \epsilon) + A_\mu(x + \epsilon)] 
\int_{x - \epsilon}^{x + \epsilon} dy^\alpha A_\alpha(y)
\nonumber \\
&& + [\int_{x - \infty}^{x - \epsilon} dy^\alpha 
\, A_\alpha(y) 
+ \int_{x - \infty}^{x + \epsilon} dy^\alpha \, A_\alpha(y)]
\, \int_{x - \epsilon}^{x + \epsilon} dy^\beta \, F_{\mu
\beta}(y) \bigg]\,.
\end{eqnarray}
It is now obvious that through the incorporation of point splitting in the gauge
transformation, one obtains a nonlocal Lagrangian density.  It is also
non-Hermitian.
If $\epsilon \rightarrow 0$ then ${\cal L}^{(1)}$ reduces to the standard QED
interaction Lagrangian density, and ${\cal L}^{(n)} \rightarrow 0$ for $n \geq 2$.\\

The appearance of infinite line integrals is an unattractive feature of this
approach, which will be remedied in Section~4.

\section{Second example}

In this example, the separation between the different space-time points is still
characterized by a fixed four-vector $\epsilon$, but for the
construction
of the infinitesimal generalized gauge transformations, one takes
an average over the separation in $ \epsilon$ using a weight function
$\rho(\eta)$.  The generalized gauge 
transformations of the fermion fields are again infinite series, the first order
term being
\begin{eqnarray}
\delta_\Lambda^{(1)} \, \psi(x) &=& - \imath \int_{- \infty}^{+ \infty}
\rho(\alpha) \, d\alpha 
\int_{- \infty}^{+ \infty} \rho(\beta) \, d\beta \int_{- \infty}^{+ \infty}
\rho(\gamma) \, d\gamma 
\nonumber \\
&&\times \, \Lambda(x + (\alpha + \gamma)\epsilon) \, \psi(x + (\beta + 
\gamma)\epsilon) \, . \label{eq11}
\end{eqnarray}
To avoid infinite line integrals, the weight function $\rho(\eta)$ is taken 
to be real and even.  It is also normalized
\begin{equation}
\int_{-\infty}^{+ \infty} d\eta \, \rho(\eta) = 1 \, ,
\end{equation} 
which guarantees that $\delta_\Lambda^{(1)} \, \psi(x)$ reduces to the
expression in Eqs.\,(\ref{eq2}) when $\epsilon \rightarrow 0$.  Finally, it must
obey the convolution property
\begin{equation}
\int_{- \infty}^{+ \infty} d\eta \, \rho(\eta) \, \rho(\xi 
- \eta) =  \rho(\xi) \,.
\end{equation}
An example of a function satisfying these four conditions is 
\begin{equation} \rho(\eta) = \frac{1}{\pi}\frac{\sin (\eta)}{\eta} \,.
\label{eq14}
\end{equation}

In momentum space, the generalized gauge 
transformations [Eq.\,(\ref{eq11})] become~:
\begin{eqnarray}
\delta_\Lambda^{(1)} \, \psi(k) &=& - \imath \int d^4k_1 
\int d^4k_2
\, \delta^{(4)}(k - k_1 - k_2) \, \Lambda(k_1) \, \psi(k_2) \, \tilde{\rho}(k_1 
\cdot
\epsilon) \nonumber \\
&& \times \, \tilde{\rho}(k_2 \cdot \epsilon) \, \tilde{\rho}((k_1 + k_2) \cdot 
\epsilon) \, , \label{eq15}
\end{eqnarray}
with $\tilde\rho(\omega)$ the Fourier transform of $\rho(\eta)$,
$\Lambda(k)$ and $\psi(k)$ being the Fourier transforms of $\Lambda(x)$
and $\psi(x)$ respectively.  The
properties of $\rho(\eta)$ translate into the following properties of
$\tilde\rho(\omega)$:
(i) $\tilde\rho(\omega)$ is real and even; 
(ii) $\tilde\rho(0) = 1$;
(iii) $\tilde{\rho}^2(\omega) = \tilde{\rho}(\omega)$,
implying that $\tilde\rho(\omega) = 0$ or $1$.
For the example in Eq.\,(\ref{eq14}), one has that 
\begin{eqnarray}
\tilde\rho(\omega) = 
\left\{ 
\begin{array}{ll}
1, &{\rm if} \: \: |\omega| < 1\\
0, &{\rm if} \: \: |\omega| \geq 1 \, .
\end{array}
\right.
\label{eq17}
\end{eqnarray}
To satisfy the Abelian group property of Section 2, one can take, e.g.,
\begin{eqnarray}
\delta_\Lambda^{(2)} \, \psi(k) &=& - \imath \int d^4k_1
\int d^4k_2 \int d^4k_3
\, \delta^{(4)}(k - k_1 - k_2 - k_3) \nonumber \\
&&\times \, \frac{A(k_1) \cdot \epsilon}{k_1 \cdot \epsilon} \, \Lambda(k_2) \,
\psi(k_3) \, [1 - \tilde{\rho}((k_2 + k_3) \cdot \epsilon)] \nonumber \\
&&\times \, \tilde{\rho}(k_1 \cdot \epsilon) \, 
\tilde{\rho}(k_2 \cdot \epsilon)\, \tilde{\rho}(k_3 \cdot \epsilon)\, 
\tilde{\rho}((k_1 + k_2 + k_3) \cdot \epsilon) \, . \label{eq18}
\end{eqnarray}
and so on for the higher order terms \cite{Olivie97}.
In spite of the appearance of $k_1 \cdot \epsilon$ in the denominator in
Eq.\,(\ref{eq18}), the
expression for $\delta_\Lambda^{(2)} \, \psi(k)$ is free of singularities for
$k_1 \cdot \epsilon \rightarrow 0$.  To see this, it suffices to observe that,
for $k_1 \cdot \epsilon \rightarrow 0$, $\tilde\rho((k_1 + k_2 + k_3) \cdot
\epsilon) \rightarrow \tilde\rho((k_2 + k_3)\cdot \epsilon)$ and that $[1 -
\tilde\rho(\omega)] \, \tilde\rho(\omega) = 0$ for all $\omega$.

Through this example, the relation between point splitting and the mitigation of
ultraviolet divergences becomes apparent.  If the function $\tilde{\rho}(\omega)$ is
taken as in Eq.\,(\ref{eq17}), then, in Eq.\,(\ref{eq15}) for 
$\delta_\Lambda^{(1)} \, \psi(k)$, the functions 
$\tilde\rho(k_1 \cdot \epsilon)$, $\tilde\rho(k_2 \cdot \epsilon)$ and 
$\tilde\rho((k_1 + k_2) \cdot \epsilon)$ cut off the high momentum 
components in the direction of $\epsilon$ of
$\Lambda(k_1)$, $\psi(k_2)$ and $\delta_\Lambda^{(1)} \, \psi(k)$ respectively.
A similar property holds for the second order term in Eq.\,(\ref{eq18}),
and it is also valid for the higher order terms.

These generalized gauge transformations reduce to the standard
gauge transformations in the limit $\epsilon \rightarrow 0$.  
Indeed, the first order term $\delta_\Lambda^{(1)} \, 
\psi(k)$ [Eq.\,(\ref{eq15})] reduces to
\begin{eqnarray}
\delta_\Lambda^{(1)} \, \psi(k) &=& - \imath \int d^4k_1 
\int d^4k_2
\, \delta^{(4)}(k - k_1 - k_2) \, \Lambda(k_1) \, \psi(k_2) \,, \label{eq40}
\end{eqnarray} 
because $\tilde{\rho}(0) = 1$.
Eq.\,(\ref{eq40}) is exactly the Fourier transform of the standard infinitesimal gauge 
transformation of Eq.\,(\ref{eq2}).  Furhermore, when $\epsilon \rightarrow 0$,
the second order term $\delta_\Lambda^{(2)} \, \psi(k)
\rightarrow 0$ in Eq.\,(\ref{eq18}), and similarly for the higher order terms
$\delta_\Lambda^{(n)} \, \psi(k)$ for $n > 2$.
Again, one can construct a generalized action, invariant under
these generalized gauge transformations, the details of which are given in
Ref.\,\cite{Olivie97}.

The advantage of the approach in this second example is that it can also be 
generalized to the case of Yang-Mills theories 
\cite{Gastmans96,Gastmans98}.  In those cases, one finds that it is absolutely necessary to average in the generalized
gauge transformations over the different arguments for the field operators.

One is still faced with a shortcoming in this approach: the momentum cut off of
the functions only occures in the direction of the four-vector $\epsilon$, and,
as a result, one cannot expect that all ultraviolet divergences are regulated.

\section{Conclusions}

We have shown that it is possible to construct generalized gauge transformations
for which the fields and the gauge parameters are taken at different space-time
points.  The separation between the different space-time points is characterized
by a fixed four-vector $\epsilon$.
Two examples of generalized gauge transformations and invariant actions are
presented here.
It appears that the generalized infinitesimal gauge
transformations of the fermion fields and the invariant actions are infinite
series in the coupling constant $e$.
In the limit $\epsilon \rightarrow 0$, they reduce to the standard 
expressions one encounters in QED.
In the second example, the conditions which the weight function has to satisfy
lead to the introduction of a cut off for the large momentum
components of the fields in the direction of $\epsilon$.  We are thus
led to consider generalized gauge transformations which 
only act on the small momentum components of the fields.

\section*{Acknowledgements}
I would like to thank Professors Raymond Gastmans and Tai Tsun Wu for sharing
their insights in this work with me.

\end{document}